\documentclass{bmcart}

\usepackage[utf8]{inputenc} 
\usepackage{bm}
\usepackage{graphicx}

\startlocaldefs
\endlocaldefs

\begin{document}

\begin{frontmatter}

\begin{fmbox}
\dochead{Research}


\title{Modeling partial lockdowns in multiplex networks using partition strategies}


\author[
   addressref={aff1, aff2},                   
   noteref={n1},                        
   email={adri.plazas@gmail.com}   
]{\inits{A.P.}\fnm{Adrià} \snm{Plazas}}
\author[
   addressref={aff1,aff2},
    noteref={n1}, 
   email={malvestio.irene@gmail.com}
]{\inits{I.M.}\fnm{Irene} \snm{Malvestio}}
\author[
   addressref={aff3},
   corref={aff3},
   email={michele.starnini@gmail.com}
]{\inits{M.S.}\fnm{Michele} \snm{Starnini}}
\author[
   addressref={aff1,aff2},
  corref = {aff2},
  email={albert.diaz@ub.edu},
]{\inits{A.D.-G.}\fnm{Albert} \snm{D\'{\i}az-Guilera}}


\address[id=aff1]{
  \orgname{Departament de F\'{\i}sica de la Mat\`eria Condensada, Universitat de Barcelona}, 
  \postcode{08028}                                
  \city{Barcelona},                              
  \cny{Catalonia, Spain}                                    
}
\address[id=aff2]{%
  \orgname{Universitat de Barcelona Institute of Complex Systems (UBICS), Universitat de Barcelona},
 \postcode{08028}                                
  \city{Barcelona},                              
  \cny{Catalonia, Spain}                                    
}

\address[id=aff3]{%
  \orgname{ISI Foundation},
 \postcode{10126}                                
  \city{Torino},                              
  \cny{Italy}                                    
}


\begin{artnotes}
\note[id=n1]{Equal contributor} 
\end{artnotes}

\end{fmbox}


\begin{abstractbox}

\begin{abstract} 
National stay-at-home orders, or lockdowns, were imposed in several countries to drastically reduce the social interactions mainly responsible for the transmission of the SARS-CoV-2 virus.
Despite being essential to slow down the COVID-19 pandemic, these containment measures are associated with an economic burden. 
In this work, we propose a network approach to model the implementation of a partial lockdown, breaking the society into disconnected components, or partitions.
Our model is composed by two main ingredients: a multiplex network representing human contacts within different contexts, formed by a Household layer, a Work layer, and a third Social layer including generic social interactions, 
and a Susceptible-Infected-Recovered process that mimics the epidemic spreading.
We compare different partition strategies, with a twofold aim: reducing the epidemic outbreak and minimizing the economic cost associated to the partial lockdown. 
We also show that the inclusion of unconstrained social interactions dramatically increases the epidemic spreading, while different kinds of restrictions on social interactions help in keeping the benefices of the network partition.
\end{abstract}




\end{abstractbox}
%

\end{frontmatter}




\section{Introduction}
The current COVID-19 pandemic left already more than one million deaths around the world and this number will increase in the near future \cite{782b1eb6f1e248bf9e8c84cf8186734f}.
As a response, almost all countries implemented unprecedented measures to restrict individual mobility and promote social distancing.
Starting by mid-March, several governments adopted a number of Non-Pharmaceutical Interventions (NPIs), whose severity rapidly increased in time: starting from school and university closures, large social gatherings avoidance, closure of non-essential activities, and finally national stay-at-home order, or lockdown \cite{Hsiang2020}.
The aim of these NPIs was to reduce and possibly interrupt the transmission of the SARS-CoV-2 virus.
National lockdowns have demonstrated very effective in
slowing down the spread of the COVID-19 epidemic \cite{e9bc4e3d4aa44d448462a15f7827be08, Walker413}, as the time-varying reproduction number of the epidemic --representing the mean number of secondary infections generated by one primary infected individual, over the course of an epidemic\cite{Liu2018}-- started to significantly drop few days after their implementation \cite{YOU2020113555,Starnini2020.06.26.20140871}.
Despite being essential to contain the pandemic, these measures deeply affected the state of the economy, triggering a major world recession \footnote{https://www.worldbank.org/en/news/press-release/2020/06/08/covid-19-to-plunge-global-economy-into-worst-recession-since-world-war-ii or a related post}.
For this reason, drastic NPIs such as national lockdowns have been adopted only for a limited time span. 
This apparent trade-off between public health and economy sparkled a heated debate regarding the optimal duration and intensity of lockdowns.

The ultimate goal of all NPIs is to decrease the number, duration, and frequency of social contacts among individuals, so to reduce the probability of virus transmission. 
The unfolding of social interactions can be represented by social networks, where nodes represent individuals and links stand for interactions \cite{Jackson2010}.
Network science has demonstrated to be a crucial tool to understand, model, and predict phenomena of social dynamics \cite{Newman2010,Castellano09}. 
The theoretical framework of network science has been recently enriched by two key concepts: Multi-layer networks~\cite{Boccaletti20141,Aleta2020a}, whose edges belong to different layers, representing different kinds of interactions; and temporal networks, whose edges appear and disappear in time, representing interactions switching on and off with given characteristic time scales~\cite{hs12,Holme2015}.
Both concepts have proved very useful for a deeper understanding of the dynamical processes on top of real networks, such as epidemic spreading \cite{burstylambiotte2013, De-Domenico:2016aa, Starnini:2017aa,estrada2020review}.

Within a network perspective, the implementation of lockdowns can be effectively seen as link removal processes, in which nodes represent individuals and links stand for their social interactions operated in different contexts, such as at work, home, or school. 
At the same time, the unfolding of the COVID-19 pandemic can be effectively represented as the spreading of an epidemic process on such networks \cite{PastorRev}.
The aim of this study is to evaluate different (partial) lockdown strategies, which consist in splitting the society into disconnected components, uniquely identified by a "color". 
In the network science domain, this translates into removing links.
Here, we assume that these links are removed with respect to social interactions occurring at workplaces, since other options (e.g. splitting households) are less feasible. 
This is associated, however, to an economic cost. 
In this work, we compare different strategies aimed at reducing such economic loss, while at the same time controlling the epidemic spread. 

The structure of the paper is the following. In Section \ref{sec:model} we describe the social network modelling the substrate responsible for the virus transmission. 
In Section \ref{sec:partitions} we propose different partition strategies, corresponding to different possible partial lockdowns. 
In Section \ref{sec:results} we show the results of numerical simulations of the epidemic spreading, while in Section \ref{sec:social} we address the effects of the inclusion of social interactions. 
Finally, Section \ref{sec:conc} is devoted to conclusions.

\section{Modelling social interactions as a multiplex network}
\label{sec:model}

The spreading of a disease like COVID-19 requires close contacts during certain time. 
Along a normal working day, people engage in different social interactions that can potentially infect others. 
These contacts are usually modelled by networks, where different activities can be represented as different layers of a multiplex network \cite{multiplex_rev}. 
For the sake of simplicity, here we consider individuals to have three kinds of interactions,
similarly to what is done in \cite{Aleta2020}, represented by three layers: Household, Work, and Social. 
According to the multiplex construction, individuals are the same across layers. 
Work and Household layers represent the strongest and well characterized fixed sets of connections in our daily life. 
A third Social layer is introduced to take into account the random and time-dependent social interactions that represent for instance shopping, using public transportation, gyms, meeting friends, and so on. 

\begin{figure}[tbp]
    \centering
    \includegraphics[width=.9\linewidth]{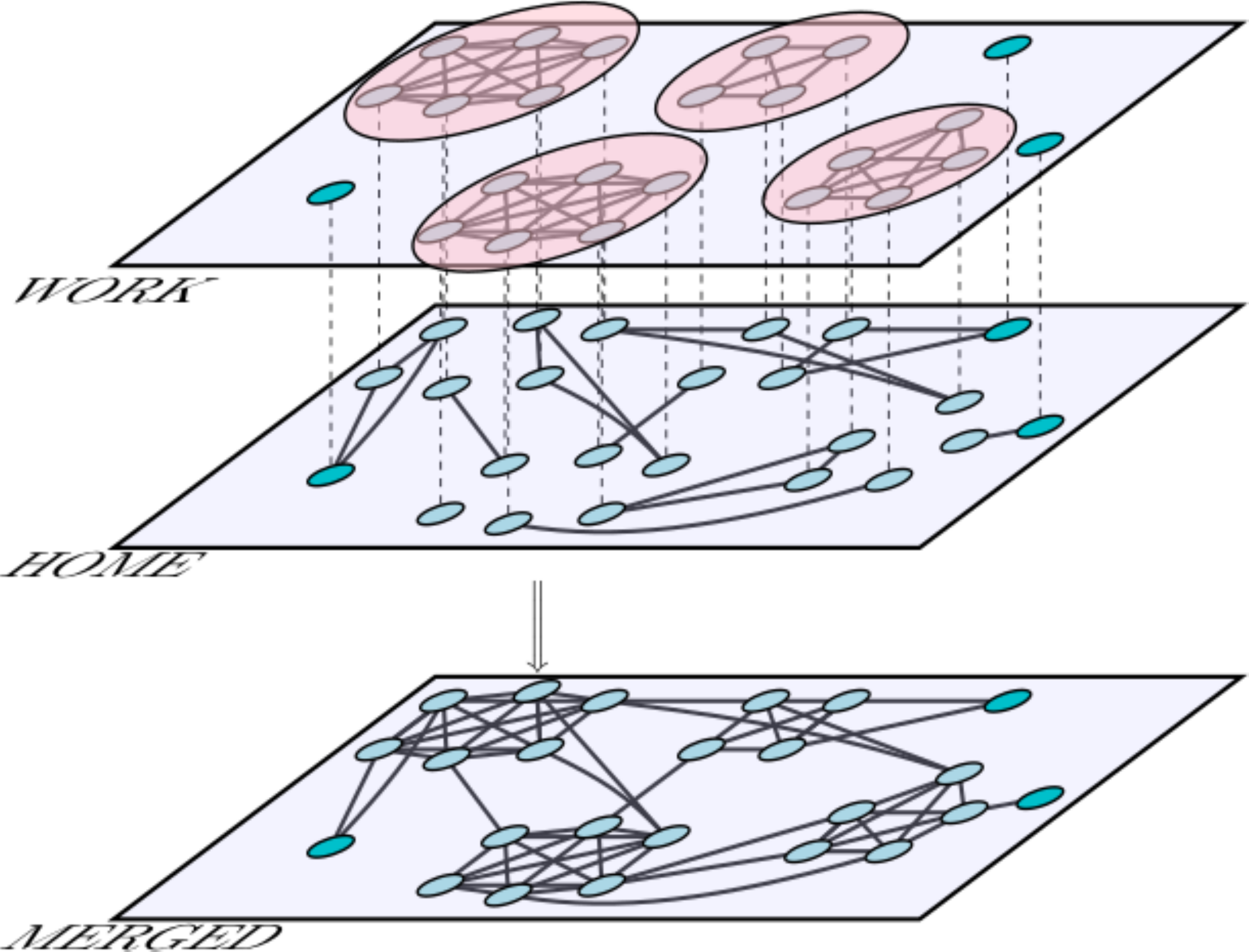}
    \caption{Multiplex construction of the two basic and fixed sets of connections. Top layer corresponds to the Work construction formed by cliques of relative large size. Bottom layer corresponds to the Household construction formed by cliques of very small size. Dashed lines are drawn just to recall that the nodes in the two layers correspond to the same individual. Merging these two layers into a single one results in a largest component, as shown in the single merged layer in the bottom of the panel.}
    \label{fig:multiplex}
\end{figure}

In this way, all individuals who work in the same company, or department (from now on, workplace) are connected in a clique (a complete subgraph) in the Work layer, which is disconnected from cliques corresponding to other companies. 
A fraction of individuals is considered to work from home (or unemployed), thus being represented as isolated nodes in the Work layer, because they do not have close contact to other working people. 
All individuals in a household form a small clique in the Household layer, which is disconnected from cliques generated by other households. 
Individuals can also interact in the Social layer, according to different hypotheses to be described in Section \ref{sec:social}. 
Although Work and Household layers are formed by isolated cliques, it is the superposition of all the layers what generates a large connected component where a disease can easily spread. This basic multiplex construction (excluding the Social layer) is visualized in Fig.~\ref{fig:multiplex}. This will be our initial setting that can be changed by containment measures.
 One does not observe a connected component in any of the layers but, if layers are merged a very large connected component (not necessarily spanning the whole network) shows up. 
 Intuitively, this makes it easier for an epidemic process to reach most of the population.
 
Here, we focus on modelling a typical urban area, in which social interactions are more frequent and disease transmission easier. 
Note that the interplay between several urban and rural areas could be modelled through the adoption of a meta-population structure \cite{COLIZZA2008450}.
We set a working scenario with the following distribution of household sizes:  1 member $0.38$, 2 : $0.38$, 3: $0.14$, 4: $0.08$, 5: $0.015$, 6: $0.005$. 
This data has been taken from the empirical distribution of the city of Barcelona \footnote{Percentages obtained from the study of the Barcelona municipal register 2019: https://www.bcn.cat/estadistica/catala/dades/tpob/llars/padro/a2019/edat/t23.htm}. 
The distribution of workplace size is assumed to be a Gaussian with mean 15 and standard deviation 5. 
Other choice for the household and workplace size distribution are shown in the supplementary material. 
We also consider that $25\%$ of the population is not part of any workplace (these people either work from home or are unemployed). 
Notice that we consider that the maximum possible number of people who work from home is included already in this group. 
In the following simulations, the size of the network is set to $N = 5000$. 

\section{Strategies for network partition} \label{sec:partitions}

The aim of this study is to evaluate different (partial) lockdown strategies, which consist in splitting the society into disconnected components, identified by a ``color". 
The goal of this procedure is to make barriers to stop the disease spreading. 
We first assign colors to the workplaces. 
As individuals live in households, it can happen that individuals of the same household have different colors assigned. This situation would produce a conflict, namely a possible path of transmission to other companies, and hence to other families and so on, if there are infected individuals. 
In order to cut these potential paths of disease transmission, we solve the conflict by removing the node(s) from the Work layer and assigning all household members the same color. 
This has, however, an associated economic cost: individuals removed from their workplace cannot work from home (since we have already included the maximum percentage of work from home in the model).
The aim of the strategies is to minimize the number of conflicts in order to reduce the economic loss but balancing it with a low disease transmission rate.

A strategy consists in assigning one  color $C_j$ to each node $j=1, \dots, N$, from a set $\{c_i$,  $i = 1, \dots, N_c \}$, where $N_c$ is the total number of colors. $N_c$ is a free parameter of the model; its limiting values are 1, corresponding to the original merged network which is the most fragile case in terms of the epidemic spread, and the number of workplaces, which is a complete segregation representing the worst economic scenario.

For the color assignment, we would ideally like to  reach two (opposite) objectives: From one side, we would like to limit the fraction of conflicts ($\chi$); from the other side, we want to make sure that the assignment is as effective as possible in limiting the spreading of the virus. 
The effectiveness of a network segregation in slowing down the spreading is related to the size distribution of its components. This is because the larger a component is the more vulnerable it is to the disease \cite{Newman2002}. 
We propose two magnitudes to quantify this vulnerability. On the one hand,  the fraction of nodes that are part of the largest connected component of the network, 
$G$. 
On the other hand, we propose the entropy of the color distribution, which measures how much the color distribution is homogeneous. 
If $p_{c_i}$ is the fraction of nodes in the network with color $c_i$, the normalized entropy $S\left( \{ p_{c_i}\}\right)$ of the color distribution is:
$$S\left( \{ p_{c_i}\}\right) = \frac{1}{\log N_c^{-1}} \sum_{i = 1}^{N_c} p_{c_i} \log p_{c_i}.$$  

We consider four strategies for the color assignment, with a fixed number of colors $N_c$.

\begin{description}
    \item[\emph{Random}] 
   We start by assigning a different color to each workplace (corresponding to a clique). 
    The number of colors is reduced by randomly merging pairs of workplaces that now will have the same color,  until we reach the desired total number of colors $N_c$. 
    Then we solve conflicts (household members with different colors) by removing nodes from the Work layer. 
    \item[\emph{Aggregation}] 
    Similar to the previous one: 
    we start with a different color for each workplace; then we iterate by merging colors together. The first color $c_i$ to be merged is selected at random, while, differently from the random case, the second color $c_j \neq c_i$ corresponds to (one of) the most popular colors among the neighbours node of $c_j$, computed over all the network. We stop when the desired number of colors $N_c$ has been reached. 
    \item[\emph{Optimized}] 
    This strategy works similarly but it aims at a more homogeneous color distribution, thus it merges first those colors assigned to the smallest set of nodes.
    The algorithm is the same as for \emph{Aggregation}, with the difference that the first color $c_k$ to be merged is not selected at random, but as the rarest between the colors available.
    The merging procedure stops once the desired number of colors $N_c$ is reached. 
    \item[\emph{Segregation}] In this strategy, instead, we start with all nodes having assigned the same color. Then, we remove nodes from the Work layer in descendent order of their  betweenness centrality \cite{freeman1977set}, up to obtaining a number of disconnected components equal to the desired number of colors $N_c$. The betweenness centrality is recomputed after each node removal. 
\end{description}

\begin{figure}[tbp]
    \centering
    \includegraphics[width=.9\linewidth]{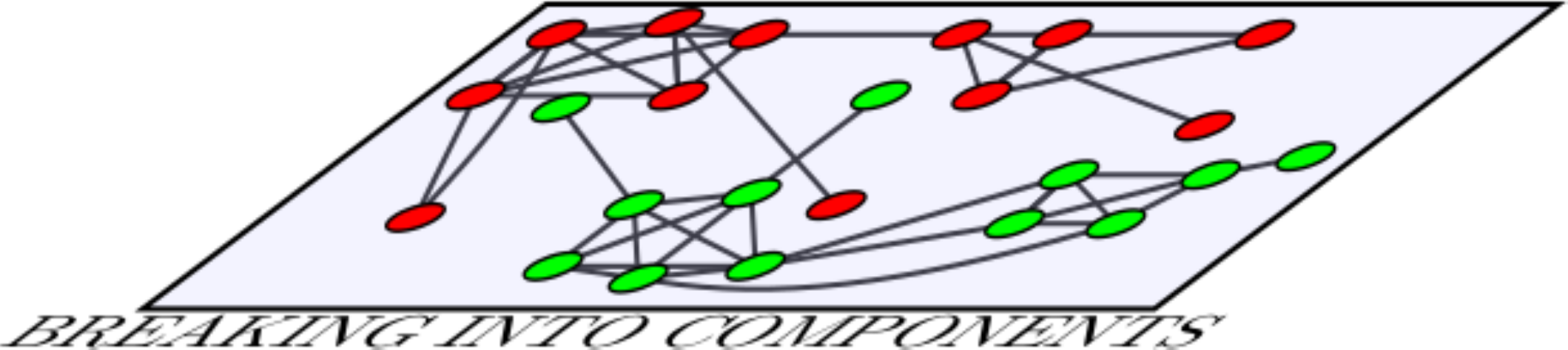}
    \caption{The merged network that gives rise to a single connected component in Fig.~\ref{fig:multiplex} is now splitted into two components (red and green) by removing some work links}
    \label{fig:colored}
\end{figure}

Fig.~\ref{fig:colored} shows the multiplex network introduced in Fig.~\ref{fig:multiplex} as broken into two independent components. 
Sizes of the components are very similar and the node removal from the Work layer has been minimized. 
Individuals who work from home have assigned the color that corresponds to their household.

\begin{figure}[tbp]
 \caption{Network of workplaces. Each node represents a different workplace, whose size is proportional to the number of people working in that workplace.
    If the workers of two workplaces are part of the same household, the workplaces are connected by a link. The link width is proportional to the number of common housemates.
     The smallest dots represent people who work from home or do not work (workplace of size $1$). 
    The nodes are colored according to the strategy (different for each panel from (a) to (d)). 
    The number of colors chosen is $N_c = 10$.
    \vspace{1cm}}
 \includegraphics[width=1.\linewidth]{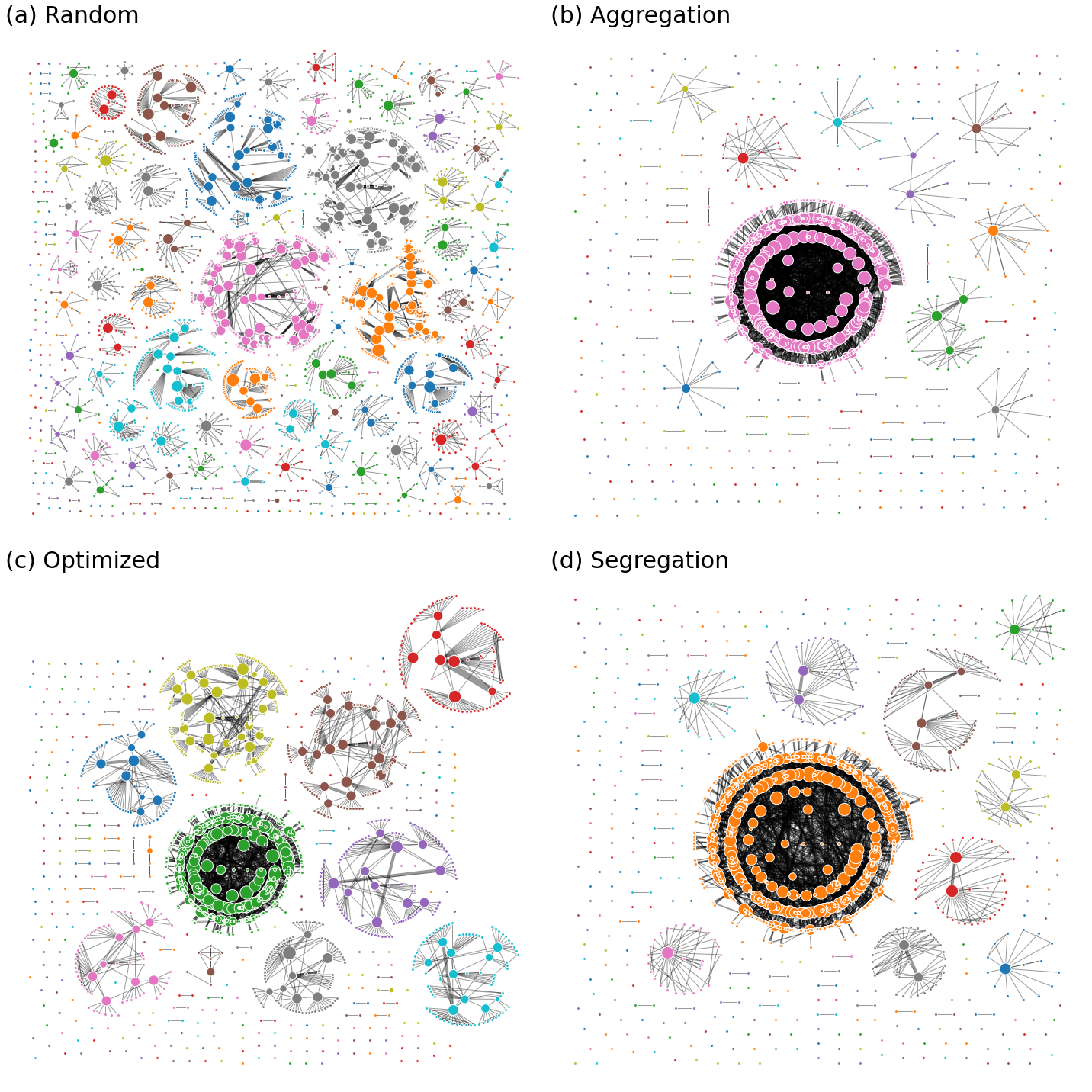}
    \label{fig:networks_colored}
\end{figure}

Fig.~\ref{fig:networks_colored} shows a representation of the network partitioned according to different strategies. 
The nodes represent the workplace cliques, or alternatively, single individuals who work from home. 
The links connect two nodes if there is at least one pair of members sharing a household.
 The colors identify separate connected components in the networks. 
 We can see how the \emph{Random} strategy has a more uniform color distribution but clearly disconnects the network into many small components, while at the other extreme \emph{Aggregation} and \emph{Segregation} strategies show the predominance of one color corresponding to one big connected cluster. 
 The \emph{Optimized} strategy is somehow a balance between these two behaviours.

\begin{figure}[tbp]
\centering
\includegraphics[width=.8\linewidth]{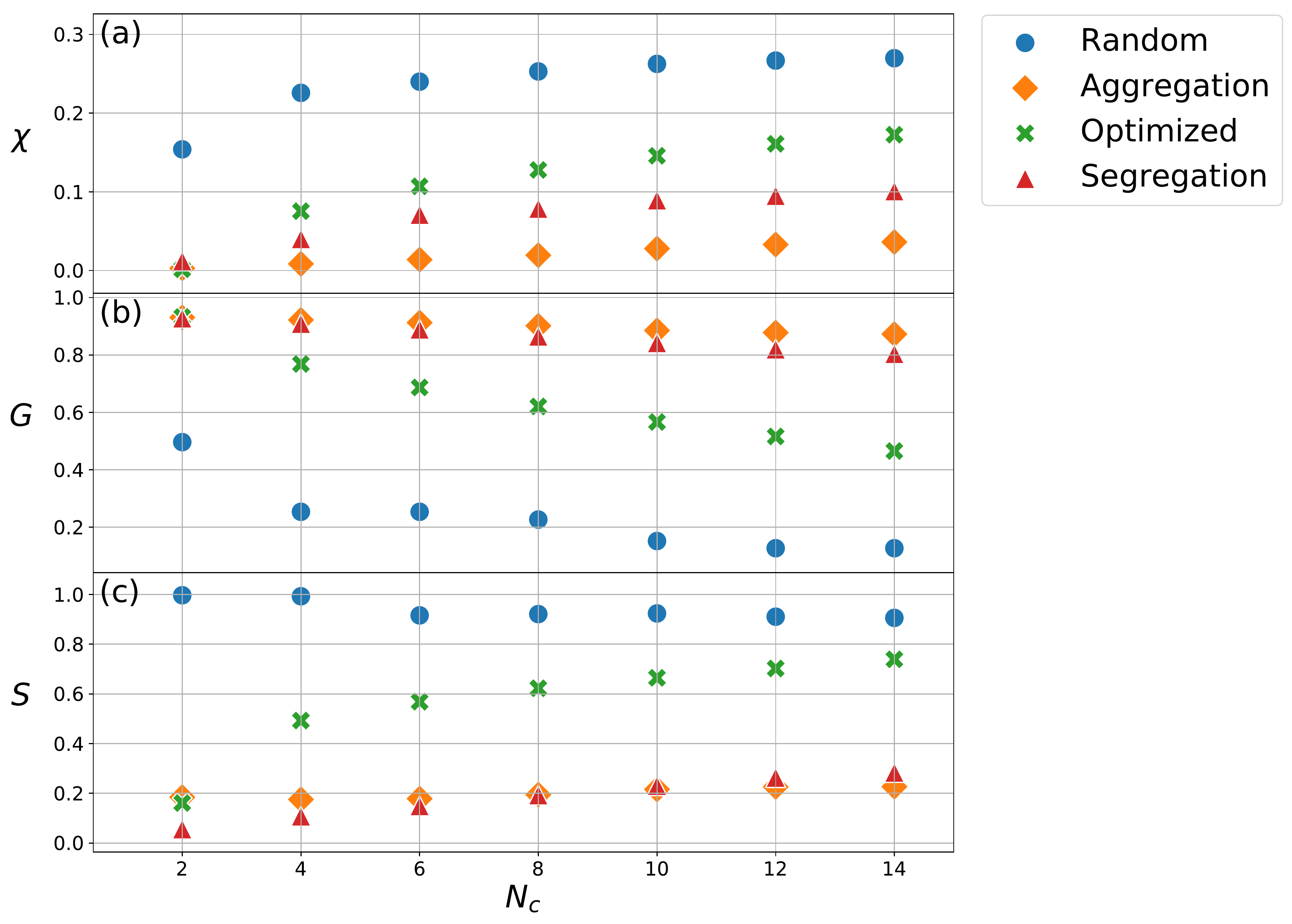}
\caption{Fraction of conflicts $\chi$ (a), relative size of the biggest connected component $G$ (b), and entropy of the color distribution $S$ (c) as a function of the number of colors in the network $N_c$, for the different containment strategies.}
\label{fig:comparison_method}
\end{figure}

In Fig.~\ref{fig:comparison_method} we compare the performance of the different strategies for the network partition. 
For each strategy, we show the fraction (with respect to the whole population) of conflicts in the network $\chi$, the relative size of the largest connected component $G$, and the entropy $S$ of the color distribution, as a function of the number of colors $N_c$.  
As expected, an increasing number of colors gives a more disconnected network (smaller $G$) and a more homogeneous color distribution (higher $S$), while the number of conflicts $\chi$ increases.
For the \emph{Random} and \emph{Aggregation} strategies, 
the entropy is almost constant for the interval of colors considered. 
Coherently to what we notice in Fig.~\ref{fig:networks_colored}, the \emph{Random} method  is the one that gives the best performance in term of breaking the connectivity of the network (smallest size of the largest connected component) and has a more homogeneous color distribution, as assessed by the entropy. 
But the \emph{Random} method is also the one with the highest percentage of conflicts, as expected. 
\emph{Aggregation} and \emph{Segregation} have the opposite trend: a low number of conflicts, but a
very high proportion of the nodes are still connected, while the color distribution is not very homogeneous. Finally, the \emph{Optimized} method can provide an intermediate number of conflicts, while being able to disrupt the network connectivity with a relatively low number of colors ($N_c \approx 10$).

 We also study the dependence of these results on the workplace and family size distribution (see of Supplementary Material Section~I, Fig.~\ref{fig:company_size} and Fig.~\ref{fig:family_size}). 
 We notice that the influence of the company size distribution is minimal.
The household size distribution, instead, has more influence for our results (Fig.~\ref{fig:company_size}):  
the inclusion  of household with more members, gives rise to a higher number of conflicts and bigger connected components.

The theoretical problem of finding a division in colors of the network that leads to the segregation of the largest connected component reminds the work of Kundu and Manna on colored percolation \cite{kundu2017colored}. In their work though the links are possible only between nodes of different colors, while in our case links are possible within the same colors. Future theoretical work can focus on the difference between these two scenarios. 

\section{Epidemic spreading} 
\label{sec:results}

In order to assess the effectiveness of the lockdown strategies presented in the previous section, we simulate the SIR (Susceptible-Infected-Recovered) model \cite{kermack1927} on the multiplex network. Briefly, in the SIR model agents are divided in the three compartments and have a probability rate to become Infected if they are Susceptible (namely the infection probability $\beta$) or to become Recovered if they are Infected (namely the recover probability $\gamma$).

As mentioned previously, our network is a multilayer formed by the Work, Household, and Social layers. In this multilayer the infection spreads following a SIR-like infection, but considering the infection probability as a node variable in order to simulate a more realistic situation that takes into account how the spread of the infection evolves differently in each layer.
We assume the infection probability $\beta$ to depend on the layer and on the node's degree \cite{Aleta2020,Perez2020}, $\beta^L_i = \frac{\beta^L}{\kappa^L_i}$. 
Here we indicate with $\beta^L_i$ the infection probability of node $i$ in layer $L$, $\beta^L$ the overall infection probability in layer $L$,  
and $\kappa^L_i$ the degree of node $i$ in layer $L$.
We then set the parameters $\beta^L$, $\gamma$ and the initial fraction of infected agents, assuming $\beta^{household}=0.50$, $\beta^{work}=0.30$, $\beta^{social}=0.20$ (so that $\sum_L \beta^L = 1$), that implies that the transmission probability is larger in households than at work. 
We also assume $\gamma=0.30$ and the initial fraction of infected agents to be equal to $1\%$.

\begin{figure}[tbp]
\centering
\includegraphics[width=.8\linewidth]{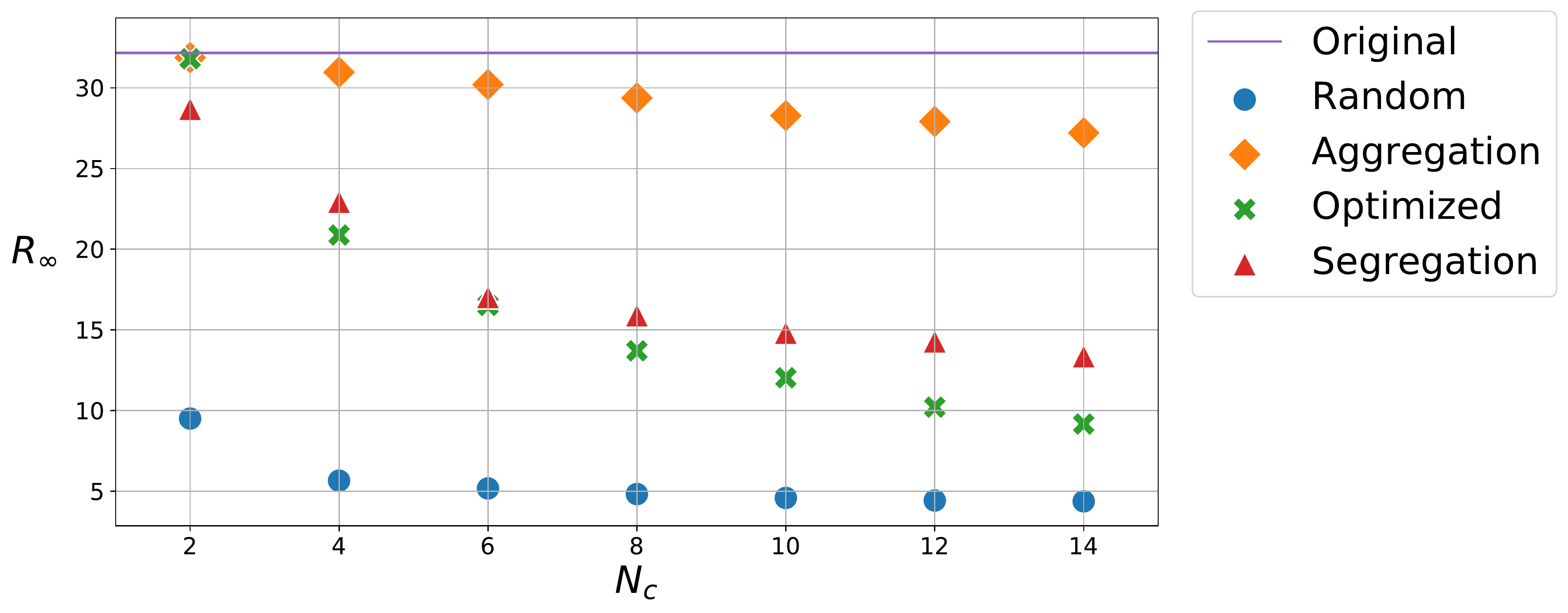}
\caption{Final fraction of recovered agents $R_{\infty}$  for the considered strategies as a function of the number of colors $N_c$. The purple line represent the corresponding value for the original case with no color. Error bars are not plotted for being negligible.}
\label{fig:removed_comparison}
\end{figure}

We start our exploration of the effect of different lockdown strategies without considering the Social layer, namely by setting $\beta^{social}=0$. 
Figure \ref{fig:removed_comparison} shows the final fraction of recovered individuals $R_\infty$, averaged over 1000 runs. 
As expected, the more colors the less infected individuals, due to a more disconnected network. 
With that in mind, the \emph{Random} strategy is the best to contain the spread of the epidemic, because of the smaller size of the largest connected component, followed by the \emph{Optimized} strategy, the \emph{Segregation} strategy and finally the \emph{Aggregation} strategy, in consonance with the results presented in Fig.~\ref{fig:comparison_method}.
Note that these strategies are ranked precisely in the reverse order with respect to the number of conflicts $\chi$ (see Fig.~\ref{fig:comparison_method}).

\section{Effect of Social interactions}
\label{sec:social}

We now explore the effects of social interactions, by adding the Social layer that accounts for all interactions that do not occur at work or in households.
This layer is built as Erdos Reny graph with average connectivity $\kappa_{mean}$. 
In other words, each node in the Social layer will randomly connect, on average, to a given number of nodes $\kappa_{mean}$ assumed to be 4. 
Moreover, we considered four different situations about how the links can be formed in the Social layer. 
Links can indeed be added by respecting color, that is, two nodes can be connected in the Social layer only if they share the same color, or without respecting it.
Furthermore, links can be generated with or without memory, that is, at every time step the links are rewired or not, respectively. 

\begin{description}
    \item[Unconstrained (no memory, no color)] the Social layer is time-dependent, changing at each time step, and links are randomly formed between nodes regardless of the color.
    \item[Color] the Social layer is time-dependent, changing at each time step, and links are randomly formed only between nodes of the same color.
    \item[Memory] the Social layer is static, without time-dependence, and links are randomly formed between nodes regardless of the color.
    \item[Fully constrained (memory and color)] the Social layer is static, without time-dependence, and links are randomly formed only between nodes of the same color.
\end{description}

\begin{figure}[tbp]
\centering
\includegraphics[width=.75\linewidth]{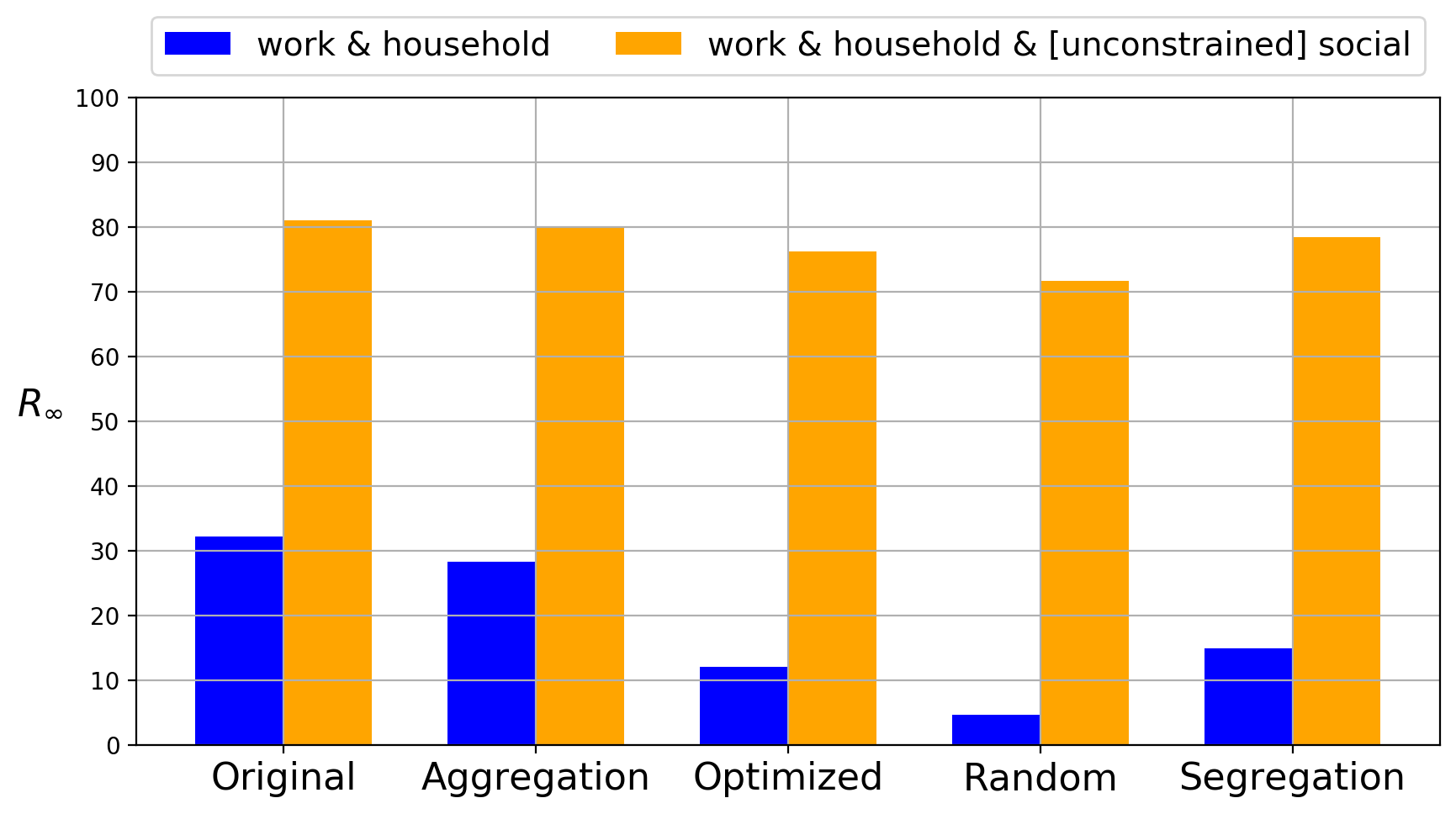}
\caption{Fraction of recovered agents, $R_{\infty}$, for different strategies. Here we compare a multiplex with only Work and Household layers (without Social layer) and a multiplex which includes also an unconstrained Social layer. The color strategies  are for the case of 10 colors. Error bars are not plotted for being negligible.}
\label{fig:removed_social}
\end{figure}

From here on, we will include the Social layer in the multiplex representation and consider the four possible forms of social interactions described above.
First, we analyze the effects of including a  Social layer whose interactions do not follow any restriction:  the unconstrained case (no memory, no color). 
In Fig.~\ref{fig:removed_social} we compare the consequences of including the Social layer in the worst scenario possible (unconstrained), compared to the original setting (without the Social layer) for the case of 10 colors. 
 Fig.~\ref{fig:removed_social} shows that with the addition of the Social layer the unconstrained case results in a much larger fraction of infected individuals, regardless of the strategy employed for the network partition. 
 We conclude that the inclusion of an unconstrained Social layer destroys the effectiveness of the colored strategies to contain the epidemic.

\begin{figure}[tbp]
\centering
\includegraphics[width=.8\linewidth]{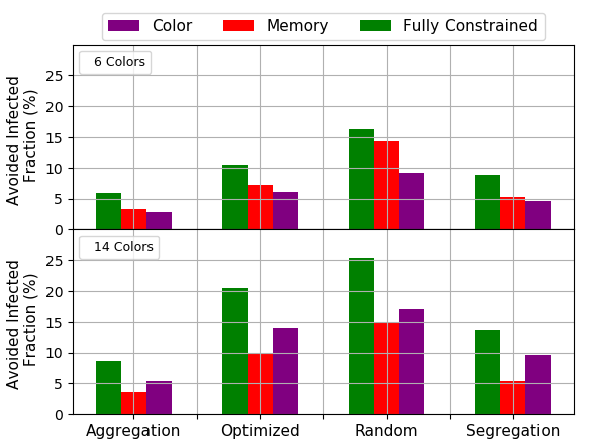}
\caption{Fraction of avoided infected agents considering different forms of social interactions compared to the scenario of no memory and not respecting the colors for the original network. Top panel: strategies are with $N_c = 6$; bottom panel: $N_c = 14$. Error bars are not plotted for being negligible.
}
\label{fig:avoided_social}
\end{figure}

In the following we explore more restricted forms of social interactions. 
Figure \ref{fig:avoided_social} shows the fraction of avoided infections for the different forms of social interactions, compared to the previously studied unconstrained situation, for a case with a small number of colors (6) and a case with a large number of colors (14). 
Several comments are in order. 
First, the greater the number of colors, the better the containment of the epidemic. Second, the best strategies are the ones that return a more disconnected network, namely the \emph{Random} and \emph{Optimized} strategy. Finally, for a small number of colors, forcing social interactions with memory (within a fixed set of individuals) is more effective than imposing  colored social interactions (with individuals sharing the same color). 
At the opposite, for a larger number of colors, it is more effective to constrain social interactions within individuals sharing the same color instead of social interactions with memory. 
In both cases,  fully constrained social interactions with both colors and memory is the most effective strategies to avoid additional infections.

In the Table~\ref{tab:comparison} we summarize the fraction of recovered individuals $R_{\infty}$ for different compositions of the multiplex network and with different constraints in building the Social layer. 
One can see that the implementation of an \emph{Optimized} partition strategy is extremely effective in reducing the impact of the disease, of about two thirds. At the same time, the economic burden of such strategy is relatively limited, with $14\%$ of individuals forced to stay home and temporarily lose their job. 
However, the inclusion of a Social layer dramatically alter this picture: the epidemic outbreak increases from $32\%$ to $81\%$ of the population. 
This is due to the fact that, in a network formed by interacting cliques, the presence of a small number of random interactions can dramatically worsen the effect of an epidemic. 
Finally, one can see that such epidemic outbreak can be reduced by imposing constraints to social interactions. 
The most effective one is the joint application of color and memory constraints in building the Social layer. 

\begin{table}[tbp]
\caption{Fraction of recovered agents $R_{\infty}$ for different compositions of the multiplex network: Work and Household only, and with the inclusion of a Social layer.  We compare the cases with no restrictions and with different combinations of constraints. 
 We consider here the \emph{Optimized} strategy with ten colors, whose fraction of conflicts is $\chi=14\%$.}
\label{tab:comparison} 
\centering
\begin{tabular}{|c|c|c|}
 \hline 
Network layers & Constraints &   $R_{\infty}$ \\
  \hline \hline 
  Work \& Household  & $-$  & $32 \%$ \\
     & Color (\emph{Optimized}) & $12 \%$ \\
        \hline
Work \& Household  \& Social &  $-$  & $81 \%$  \\
 & Memory &  $78 \%$ \\
 & Color (\emph{Optimized}) & $71\%$  \\
 & Color (\emph{Optimized}) + Memory  & $64 \%$  \\
  \hline 
 \end{tabular}
\end{table}

\section{Conclusions}
\label{sec:conc}

In this work we proposed a network approach to model the implementation of different strategies for a partial lockdown. 
Our model is composed by two main ingredients: a multiplex network including social interactions within different contexts, and numerical simulations of a SIR process to mimic the epidemic spreading. 
We proposed different strategies to segregate the network into disconnected components (partitions) with a twofold goal:
halting the epidemic spreading, whose effectiveness can be measured by the reduction in the number of infected individuals, and minimizing the economic burden of the partial lockdown, that can be quantified by the removed links in the Work layer, that represent job losses. 
We found that the best partition strategy for containing the epidemic is a \emph{Random} strategy, but this comes with the larger job loss. A good compromise is the so-called \emph{Optimized} strategy, which is able to create a good segregation in the network while also minimizing the link removal. 
We showed that the inclusion of unconstrained social interactions dramatically increased the spreading of the disease.  As a consequences, we studied different constraints to be applied specifically to the links in the  Social layer: only within the same partition (joining nodes with the same color) and/or with memory (individuals interact with the same peers over time). 
With a number of color high enough, imposing color on the social interactions is more effective than maintaining memory, while clearly the two methods together would be the best combinations to reduce the epidemic outbreak.

Our work comes with limitations. 
For instance, the multiplex network that models social interactions is built on several assumptions and not directly by using empirical data regarding contact matrix within work or household contexts. 
Note that we did not include schools in the network modelling, assuming that schools can stay temporarily closed during partial lockdowns. 
However, data such as the precise composition of households or workplaces are not readily available. 
Furthermore, we adopted a very simple model for the epidemic spreading -- the SIR model --, which is not realistic. 
This choice is however motivated to keep the number of parameters of the model, which is quite large, as low as possible. 
In future works, it would be interesting to explore the effects of the proposed partition strategies and restriction for social interactions on empirical data, regarding both the network reconstruction and the disease propagation.
Finally, it would be interesting to include mobility data to test if the proposed partition strategies could be realistically implemented.


\section*{List of abbreviations} 
SIR: Susceptible-Infected-Recovered \\
NPIs: Non-Pharmaceutical Interventions 

\section*{Declarations}
\begin{backmatter}
\section*{Availability of data and materials}
The data that support the findings of this study are available from the corresponding authors upon reasonable request.

\section*{Competing interests}
  The authors declare that they have no competing interests.

\section*{Funding}
 A.D.-G. and I.M.  acknowledge financial support from Spanish Government via Project No. PGC2018-094754-B-C22 (MCIU/AEI/FEDER,UE) and Generalitat de Catalunya via Grant No. 2017SGR341. 
 
 \section*{Author's contributions}
   A.D.-G. designed the experiments, A.P. and I.M. performed simulations. All authors discussed the methods and results, and wrote the manuscript.
   
\section*{Acknowledgements}
 A.D.-G. is indebted to COVIDWarriors association, and specially to Jordi Bosch, for suggesting {\em PARCHIS} as a strategy
to maximize economic reactivation while minimizing the risk of infection.


\bibliographystyle{bmc-mathphys} 


\newcommand{\BMCxmlcomment}[1]{}

\BMCxmlcomment{

<refgrp>

<bibl id="B1">
  <title><p>An interactive web-based dashboard to track COVID-19 in real
  time</p></title>
  <aug>
    <au><snm>Dong</snm><fnm>E</fnm></au>
    <au><snm>Du</snm><fnm>H</fnm></au>
    <au><snm>Gardner</snm><fnm>L</fnm></au>
  </aug>
  <source>The Lancet Infectious Diseases</source>
  <pubdate>2020</pubdate>
  <volume>20</volume>
  <issue>5</issue>
</bibl>

<bibl id="B2">
  <title><p>{The effect of large-scale anti-contagion policies on the COVID-19
  pandemic}</p></title>
  <aug>
    <au><snm>Hsiang</snm><fnm>S</fnm></au>
    <au><snm>Allen</snm><fnm>D</fnm></au>
    <au><snm>Annan Phan</snm><fnm>S</fnm></au>
    <au><snm>Bell</snm><fnm>K</fnm></au>
    <au><snm>Bolliger</snm><fnm>I</fnm></au>
    <au><snm>Chong</snm><fnm>T</fnm></au>
    <au><snm>Druckenmiller</snm><fnm>H</fnm></au>
    <au><snm>Huang</snm><fnm>LY</fnm></au>
    <au><snm>Hultgren</snm><fnm>A</fnm></au>
    <au><snm>Krasovich</snm><fnm>E</fnm></au>
    <au><snm>Lau</snm><fnm>P</fnm></au>
    <au><snm>Lee</snm><fnm>J</fnm></au>
    <au><snm>Rolf</snm><fnm>E</fnm></au>
    <au><snm>Tseng</snm><fnm>J</fnm></au>
    <au><snm>Wu</snm><fnm>T</fnm></au>
  </aug>
  <source>Nature</source>
  <publisher>Springer US</publisher>
  <pubdate>2020</pubdate>
  <volume>584</volume>
  <issue>March</issue>
  <url>http://dx.doi.org/10.1038/s41586-020-2404-8</url>
</bibl>

<bibl id="B3">
  <title><p>Estimating the effects of non-pharmaceutical interventions on
  COVID-19 in Europe</p></title>
  <aug>
    <au><snm>Flaxman</snm><fnm>S</fnm></au>
    <au><snm>Mishra</snm><fnm>S</fnm></au>
    <au><snm>Gandy</snm><fnm>A</fnm></au>
    <au><snm>Unwin</snm><fnm>{H. Juliette T.}</fnm></au>
    <au><snm>Mellan</snm><fnm>{Thomas A.}</fnm></au>
    <au><snm>Coupland</snm><fnm>H</fnm></au>
    <au><snm>Whittaker</snm><fnm>C</fnm></au>
    <au><snm>Zhu</snm><fnm>H</fnm></au>
    <au><snm>Berah</snm><fnm>T</fnm></au>
    <au><snm>Eaton</snm><fnm>{Jeffrey W.}</fnm></au>
    <au><snm>Monod</snm><fnm>M</fnm></au>
    <au><snm>Perez Guzman</snm><fnm>{Pablo N.}</fnm></au>
    <au><snm>Schmit</snm><fnm>N</fnm></au>
    <au><snm>Cilloni</snm><fnm>L</fnm></au>
    <au><snm>Ainslie</snm><fnm>{Kylie E.C.}</fnm></au>
    <au><snm>Baguelin</snm><fnm>M</fnm></au>
    <au><snm>Boonyasiri</snm><fnm>A</fnm></au>
    <au><snm>Boyd</snm><fnm>O</fnm></au>
    <au><snm>Cattarino</snm><fnm>L</fnm></au>
    <au><snm>Cooper</snm><fnm>{Laura V.}</fnm></au>
    <au><snm>Cucunub{\'a}</snm><fnm>Z</fnm></au>
    <au><snm>Cuomo Dannenburg</snm><fnm>G</fnm></au>
    <au><snm>Dighe</snm><fnm>A</fnm></au>
    <au><snm>Djaafara</snm><fnm>B</fnm></au>
    <au><snm>Dorigatti</snm><fnm>I</fnm></au>
    <au><snm>{van Elsland}</snm><fnm>{Sabine L.}</fnm></au>
    <au><snm>FitzJohn</snm><fnm>{Richard G.}</fnm></au>
    <au><snm>Gaythorpe</snm><fnm>{Katy A.M.}</fnm></au>
    <au><snm>Geidelberg</snm><fnm>L</fnm></au>
    <au><snm>Grassly</snm><fnm>{Nicholas C.}</fnm></au>
    <au><snm>Green</snm><fnm>{William D.}</fnm></au>
    <au><snm>Hallett</snm><fnm>T</fnm></au>
    <au><snm>Hamlet</snm><fnm>A</fnm></au>
    <au><snm>Hinsley</snm><fnm>W</fnm></au>
    <au><snm>Jeffrey</snm><fnm>B</fnm></au>
    <au><snm>Knock</snm><fnm>E</fnm></au>
    <au><snm>Laydon</snm><fnm>{Daniel J.}</fnm></au>
    <au><snm>Nedjati Gilani</snm><fnm>G</fnm></au>
    <au><snm>Nouvellet</snm><fnm>P</fnm></au>
    <au><snm>Parag</snm><fnm>{Kris V.}</fnm></au>
    <au><snm>Siveroni</snm><fnm>I</fnm></au>
    <au><snm>Thompson</snm><fnm>{Hayley A.}</fnm></au>
    <au><snm>Verity</snm><fnm>R</fnm></au>
    <au><snm>Volz</snm><fnm>E</fnm></au>
    <au><snm>Walters</snm><fnm>{Caroline E.}</fnm></au>
    <au><snm>Wang</snm><fnm>H</fnm></au>
    <au><snm>Wang</snm><fnm>Y</fnm></au>
    <au><snm>Watson</snm><fnm>{Oliver J.}</fnm></au>
    <au><snm>Winskill</snm><fnm>P</fnm></au>
    <au><snm>Xi</snm><fnm>X</fnm></au>
    <au><snm>Walker</snm><fnm>{Patrick G.T.}</fnm></au>
    <au><snm>Ghani</snm><fnm>{Azra C.}</fnm></au>
    <au><snm>Donnelly</snm><fnm>{Christl A.}</fnm></au>
    <au><snm>Riley</snm><fnm>S</fnm></au>
    <au><snm>Vollmer</snm><fnm>{Michaela A.C.}</fnm></au>
    <au><snm>Ferguson</snm><fnm>{Neil M.}</fnm></au>
    <au><snm>Okell</snm><fnm>{Lucy C.}</fnm></au>
    <au><snm>Bhatt</snm><fnm>S</fnm></au>
  </aug>
  <source>Nature</source>
  <pubdate>2020</pubdate>
  <volume>584</volume>
  <issue>7820</issue>
</bibl>

<bibl id="B4">
  <title><p>The impact of COVID-19 and strategies for mitigation and
  suppression in low- and middle-income countries</p></title>
  <aug>
    <au><snm>Walker</snm><fnm>PGT</fnm></au>
    <au><snm>Whittaker</snm><fnm>C</fnm></au>
    <au><snm>Watson</snm><fnm>OJ</fnm></au>
    <au><snm>Baguelin</snm><fnm>M</fnm></au>
    <au><snm>Winskill</snm><fnm>P</fnm></au>
    <au><snm>Hamlet</snm><fnm>A</fnm></au>
    <au><snm>Djafaara</snm><fnm>BA</fnm></au>
    <au><snm>Cucunub{\'a}</snm><fnm>Z</fnm></au>
    <au><snm>Olivera Mesa</snm><fnm>D</fnm></au>
    <au><snm>Green</snm><fnm>W</fnm></au>
    <au><snm>Thompson</snm><fnm>H</fnm></au>
    <au><snm>Nayagam</snm><fnm>S</fnm></au>
    <au><snm>Ainslie</snm><fnm>KEC</fnm></au>
    <au><snm>Bhatia</snm><fnm>S</fnm></au>
    <au><snm>Bhatt</snm><fnm>S</fnm></au>
    <au><snm>Boonyasiri</snm><fnm>A</fnm></au>
    <au><snm>Boyd</snm><fnm>O</fnm></au>
    <au><snm>Brazeau</snm><fnm>NF</fnm></au>
    <au><snm>Cattarino</snm><fnm>L</fnm></au>
    <au><snm>Cuomo Dannenburg</snm><fnm>G</fnm></au>
    <au><snm>Dighe</snm><fnm>A</fnm></au>
    <au><snm>Donnelly</snm><fnm>CA</fnm></au>
    <au><snm>Dorigatti</snm><fnm>I</fnm></au>
    <au><snm>Elsland</snm><fnm>SL</fnm></au>
    <au><snm>FitzJohn</snm><fnm>R</fnm></au>
    <au><snm>Fu</snm><fnm>H</fnm></au>
    <au><snm>Gaythorpe</snm><fnm>KAM</fnm></au>
    <au><snm>Geidelberg</snm><fnm>L</fnm></au>
    <au><snm>Grassly</snm><fnm>N</fnm></au>
    <au><snm>Haw</snm><fnm>D</fnm></au>
    <au><snm>Hayes</snm><fnm>S</fnm></au>
    <au><snm>Hinsley</snm><fnm>W</fnm></au>
    <au><snm>Imai</snm><fnm>N</fnm></au>
    <au><snm>Jorgensen</snm><fnm>D</fnm></au>
    <au><snm>Knock</snm><fnm>E</fnm></au>
    <au><snm>Laydon</snm><fnm>D</fnm></au>
    <au><snm>Mishra</snm><fnm>S</fnm></au>
    <au><snm>Nedjati Gilani</snm><fnm>G</fnm></au>
    <au><snm>Okell</snm><fnm>LC</fnm></au>
    <au><snm>Unwin</snm><fnm>HJ</fnm></au>
    <au><snm>Verity</snm><fnm>R</fnm></au>
    <au><snm>Vollmer</snm><fnm>M</fnm></au>
    <au><snm>Walters</snm><fnm>CE</fnm></au>
    <au><snm>Wang</snm><fnm>H</fnm></au>
    <au><snm>Wang</snm><fnm>Y</fnm></au>
    <au><snm>Xi</snm><fnm>X</fnm></au>
    <au><snm>Lalloo</snm><fnm>DG</fnm></au>
    <au><snm>Ferguson</snm><fnm>NM</fnm></au>
    <au><snm>Ghani</snm><fnm>AC</fnm></au>
  </aug>
  <source>Science</source>
  <publisher>American Association for the Advancement of Science</publisher>
  <pubdate>2020</pubdate>
  <volume>369</volume>
  <issue>6502</issue>
  <fpage>413</fpage>
  <lpage>-422</lpage>
  <url>https://science.sciencemag.org/content/369/6502/413</url>
</bibl>

<bibl id="B5">
  <title><p>{Measurability of the epidemic reproduction number in data-driven
  contact networks}</p></title>
  <aug>
    <au><snm>Liu</snm><fnm>QH</fnm></au>
    <au><snm>Ajelli</snm><fnm>M</fnm></au>
    <au><snm>Aleta</snm><fnm>A</fnm></au>
    <au><snm>Merler</snm><fnm>S</fnm></au>
    <au><snm>Moreno</snm><fnm>Y</fnm></au>
    <au><snm>Vespignani</snm><fnm>A</fnm></au>
  </aug>
  <source>Proceedings of the National Academy of Sciences of the United States
  of America</source>
  <pubdate>2018</pubdate>
  <volume>115</volume>
  <issue>50</issue>
  <fpage>12680</fpage>
  <lpage>-12685</lpage>
</bibl>

<bibl id="B6">
  <title><p>Estimation of the time-varying reproduction number of COVID-19
  outbreak in China</p></title>
  <aug>
    <au><snm>You</snm><fnm>C</fnm></au>
    <au><snm>Deng</snm><fnm>Y</fnm></au>
    <au><snm>Hu</snm><fnm>W</fnm></au>
    <au><snm>Sun</snm><fnm>J</fnm></au>
    <au><snm>Lin</snm><fnm>Q</fnm></au>
    <au><snm>Zhou</snm><fnm>F</fnm></au>
    <au><snm>Pang</snm><fnm>CH</fnm></au>
    <au><snm>Zhang</snm><fnm>Y</fnm></au>
    <au><snm>Chen</snm><fnm>Z</fnm></au>
    <au><snm>Zhou</snm><fnm>XH</fnm></au>
  </aug>
  <source>International Journal of Hygiene and Environmental Health</source>
  <pubdate>2020</pubdate>
  <volume>228</volume>
  <fpage>113555</fpage>
  <url>http://www.sciencedirect.com/science/article/pii/S1438463920302133</url>
</bibl>

<bibl id="B7">
  <title><p>Impact of the accuracy of case-based surveillance data on the
  estimation of time-varying reproduction numbers</p></title>
  <aug>
    <au><snm>Starnini</snm><fnm>M</fnm></au>
    <au><snm>Aleta</snm><fnm>A</fnm></au>
    <au><snm>Tizzoni</snm><fnm>M</fnm></au>
    <au><snm>Moreno</snm><fnm>Y</fnm></au>
  </aug>
  <source>medRxiv</source>
  <publisher>Cold Spring Harbor Laboratory Press</publisher>
  <pubdate>2020</pubdate>
  <url>https://www.medrxiv.org/content/early/2020/06/28/2020.06.26.20140871</url>
</bibl>

<bibl id="B8">
  <title><p>Social and Economic Networks</p></title>
  <aug>
    <au><snm>Jackson</snm><fnm>M.</fnm></au>
  </aug>
  <publisher>Princeton: Princeton University Press</publisher>
  <pubdate>2010</pubdate>
</bibl>

<bibl id="B9">
  <title><p>Networks: An introduction</p></title>
  <aug>
    <au><snm>Newman</snm><fnm>M. E. J.</fnm></au>
  </aug>
  <publisher>Oxford: Oxford University Press</publisher>
  <pubdate>2010</pubdate>
</bibl>

<bibl id="B10">
  <title><p>Statistical physics of social dynamics</p></title>
  <aug>
    <au><snm>Castellano</snm><fnm>C.</fnm></au>
    <au><snm>Fortunato</snm><fnm>S.</fnm></au>
    <au><snm>Loreto</snm><fnm>V.</fnm></au>
  </aug>
  <source>Rev. Mod. Phys.</source>
  <pubdate>2009</pubdate>
  <volume>81</volume>
  <fpage>591</fpage>
  <lpage>-646</lpage>
</bibl>

<bibl id="B11">
  <title><p>The structure and dynamics of multilayer networks</p></title>
  <aug>
    <au><snm>Boccaletti</snm><fnm>S.</fnm></au>
    <au><snm>Bianconi</snm><fnm>G.</fnm></au>
    <au><snm>Criado</snm><fnm>R.</fnm></au>
    <au><snm>Genio</snm><fnm>C.I.</fnm></au>
    <au><snm>G{\'o}mez Garde{\~n}es</snm><fnm>J.</fnm></au>
    <au><snm>Romance</snm><fnm>M.</fnm></au>
    <au><snm>Sendi{\~n}a Nadal</snm><fnm>I.</fnm></au>
    <au><snm>Wang</snm><fnm>Z.</fnm></au>
    <au><snm>Zanin</snm><fnm>M.</fnm></au>
  </aug>
  <source>Physics Reports</source>
  <pubdate>2014</pubdate>
  <volume>544</volume>
  <issue>1</issue>
  <fpage>1</fpage>
  <lpage>122</lpage>
</bibl>

<bibl id="B12">
  <title><p>{Data-driven contact structures: From homogeneous mixing to
  multilayer networks}</p></title>
  <aug>
    <au><snm>Aleta</snm><fnm>A</fnm></au>
    <au><snm>{Ferraz de Arruda}</snm><fnm>G</fnm></au>
    <au><snm>Moreno</snm><fnm>Y</fnm></au>
  </aug>
  <source>PLoS computational biology</source>
  <pubdate>2020</pubdate>
  <volume>16</volume>
  <issue>7</issue>
  <fpage>e1008035</fpage>
  <url>http://dx.doi.org/10.1371/journal.pcbi.1008035</url>
</bibl>

<bibl id="B13">
  <title><p>Temporal networks</p></title>
  <aug>
    <au><snm>Holme</snm><fnm>P</fnm></au>
    <au><snm>Saramäki</snm><fnm>J</fnm></au>
  </aug>
  <source>Physics Reports</source>
  <pubdate>2012</pubdate>
  <volume>519</volume>
  <issue>3</issue>
  <fpage>97</fpage>
  <lpage>125</lpage>
  <url>http://www.sciencedirect.com/science/article/pii/S0370157312000841</url>
</bibl>

<bibl id="B14">
  <title><p>{Modern temporal network theory: a colloquium}</p></title>
  <aug>
    <au><snm>Holme</snm><fnm>P</fnm></au>
  </aug>
  <source>Eur. Phys. J. B</source>
  <pubdate>2015</pubdate>
  <volume>88</volume>
  <issue>9</issue>
  <fpage>234</fpage>
</bibl>

<bibl id="B15">
  <title><p>Burstiness and spreading on temporal networks</p></title>
  <aug>
    <au><snm>Lambiotte</snm><fnm>R</fnm></au>
    <au><snm>Tabourier</snm><fnm>L</fnm></au>
    <au><snm>Delvenne</snm><fnm>JC</fnm></au>
  </aug>
  <source>Eur. Phys. J. B</source>
  <pubdate>2013</pubdate>
  <volume>86</volume>
  <fpage>320</fpage>
</bibl>

<bibl id="B16">
  <title><p>The physics of spreading processes in multilayer
  networks</p></title>
  <aug>
    <au><snm>De Domenico</snm><fnm>M</fnm></au>
    <au><snm>Granell</snm><fnm>C</fnm></au>
    <au><snm>Porter</snm><fnm>MA</fnm></au>
    <au><snm>Arenas</snm><fnm>A</fnm></au>
  </aug>
  <source>Nat Phys</source>
  <publisher>Nature Publishing Group</publisher>
  <pubdate>2016</pubdate>
  <volume>12</volume>
  <fpage>901</fpage>
  <lpage>-906</lpage>
</bibl>

<bibl id="B17">
  <title><p>Effects of temporal correlations in social multiplex
  networks</p></title>
  <aug>
    <au><snm>Starnini</snm><fnm>M</fnm></au>
    <au><snm>Baronchelli</snm><fnm>A</fnm></au>
    <au><snm>Pastor Satorras</snm><fnm>R</fnm></au>
  </aug>
  <source>Scientific Reports</source>
  <pubdate>2017</pubdate>
  <volume>7</volume>
  <issue>1</issue>
  <fpage>8597</fpage>
  <url>https://doi.org/10.1038/s41598-017-07591-0</url>
</bibl>

<bibl id="B18">
  <title><p>COVID-19 and SARS-CoV-2. Modeling the present, looking at the
  future</p></title>
  <aug>
    <au><snm>Estrada</snm><fnm>E</fnm></au>
  </aug>
  <source>Physics Reports</source>
  <pubdate>2020</pubdate>
  <volume>869</volume>
  <fpage>1</fpage>
  <lpage>51</lpage>
  <url>http://www.sciencedirect.com/science/article/pii/S0370157320302544</url>
</bibl>

<bibl id="B19">
  <title><p>Epidemic processes in complex networks</p></title>
  <aug>
    <au><snm>Pastor Satorras</snm><fnm>R</fnm></au>
    <au><snm>Castellano</snm><fnm>C</fnm></au>
    <au><snm>Van Mieghem</snm><fnm>P</fnm></au>
    <au><snm>Vespignani</snm><fnm>A</fnm></au>
  </aug>
  <source>Rev. Mod. Phys.</source>
  <publisher>APS</publisher>
  <pubdate>2015</pubdate>
  <volume>87</volume>
  <issue>3</issue>
  <fpage>925</fpage>
</bibl>

<bibl id="B20">
  <title><p>{Multilayer networks}</p></title>
  <aug>
    <au><snm>Kivel{\"a}</snm><fnm>M</fnm></au>
    <au><snm>Arenas</snm><fnm>A</fnm></au>
    <au><snm>Barthelemy</snm><fnm>M</fnm></au>
    <au><snm>Gleeson</snm><fnm>JP</fnm></au>
    <au><snm>Moreno</snm><fnm>Y</fnm></au>
    <au><snm>Porter</snm><fnm>MA</fnm></au>
  </aug>
  <source>Journal of Complex Networks</source>
  <pubdate>2014</pubdate>
  <volume>2</volume>
  <issue>3</issue>
  <fpage>203</fpage>
  <lpage>271</lpage>
  <url>https://doi.org/10.1093/comnet/cnu016</url>
</bibl>

<bibl id="B21">
  <title><p>{Modelling the impact of testing, contact tracing and household
  quarantine on second waves of COVID-19}</p></title>
  <aug>
    <au><snm>Aleta</snm><fnm>A</fnm></au>
    <au><snm>Mart{\'{i}}n Corral</snm><fnm>D</fnm></au>
    <au><snm>{Pastore y Piontti}</snm><fnm>A</fnm></au>
    <au><snm>Ajelli</snm><fnm>M</fnm></au>
    <au><snm>Litvinova</snm><fnm>M</fnm></au>
    <au><snm>Chinazzi</snm><fnm>M</fnm></au>
    <au><snm>Dean</snm><fnm>NE</fnm></au>
    <au><snm>Halloran</snm><fnm>ME</fnm></au>
    <au><snm>Longini</snm><fnm>IM</fnm></au>
    <au><snm>Merler</snm><fnm>S</fnm></au>
    <au><snm>Pentland</snm><fnm>A</fnm></au>
    <au><snm>Vespignani</snm><fnm>A</fnm></au>
    <au><snm>Moro</snm><fnm>E</fnm></au>
    <au><snm>Moreno</snm><fnm>Y</fnm></au>
  </aug>
  <source>Nature Human Behaviour</source>
  <pubdate>2020</pubdate>
</bibl>

<bibl id="B22">
  <title><p>Epidemic modeling in metapopulation systems with heterogeneous
  coupling pattern: Theory and simulations</p></title>
  <aug>
    <au><snm>Colizza</snm><fnm>V</fnm></au>
    <au><snm>Vespignani</snm><fnm>A</fnm></au>
  </aug>
  <source>Journal of Theoretical Biology</source>
  <pubdate>2008</pubdate>
  <volume>251</volume>
  <issue>3</issue>
  <fpage>450</fpage>
  <lpage>467</lpage>
  <url>http://www.sciencedirect.com/science/article/pii/S0022519307005991</url>
</bibl>

<bibl id="B23">
  <title><p>{Spread of epidemic disease on networks}</p></title>
  <aug>
    <au><snm>Newman</snm><fnm>M. E.J.</fnm></au>
  </aug>
  <source>Physical Review E - Statistical Physics, Plasmas, Fluids, and Related
  Interdisciplinary Topics</source>
  <pubdate>2002</pubdate>
  <volume>66</volume>
  <issue>1</issue>
  <fpage>1</fpage>
  <lpage>-11</lpage>
</bibl>

<bibl id="B24">
  <title><p>A set of measures of centrality based on betweenness</p></title>
  <aug>
    <au><snm>Freeman</snm><fnm>LC</fnm></au>
  </aug>
  <source>Sociometry</source>
  <publisher>JSTOR</publisher>
  <pubdate>1977</pubdate>
  <fpage>35</fpage>
  <lpage>-41</lpage>
</bibl>

<bibl id="B25">
  <title><p>Colored percolation</p></title>
  <aug>
    <au><snm>Kundu</snm><fnm>S</fnm></au>
    <au><snm>Manna</snm><fnm>SS</fnm></au>
  </aug>
  <source>Physical Review E</source>
  <publisher>APS</publisher>
  <pubdate>2017</pubdate>
  <volume>95</volume>
  <issue>5</issue>
  <fpage>052124</fpage>
</bibl>

<bibl id="B26">
  <title><p>A Contribution to the Mathematical Theory of Epidemics</p></title>
  <aug>
    <au><snm>Kermack</snm><fnm>W. O.</fnm></au>
    <au><snm>McKendric</snm><fnm>A. G.</fnm></au>
  </aug>
  <source>Proceedings of the Royal Society of London</source>
  <pubdate>1927</pubdate>
  <volume>115</volume>
  <fpage>700</fpage>
  <lpage>721</lpage>
</bibl>

<bibl id="B27">
  <title><p>{Disease spreading with social distancing: a prevention strategy in
  disordered multiplex networks}</p></title>
  <aug>
    <au><snm>Perez</snm><fnm>I. A.</fnm></au>
    <au><snm>{Di Muro}</snm><fnm>M. A.</fnm></au>
    <au><snm>{La Rocca}</snm><fnm>C. E.</fnm></au>
    <au><snm>Braunstein</snm><fnm>L. A.</fnm></au>
  </aug>
  <source>Physical Review E</source>
  <publisher>American Physical Society</publisher>
  <pubdate>2020</pubdate>
  <volume>022310</volume>
  <issue>7600</issue>
  <fpage>1</fpage>
  <lpage>-22</lpage>
  <url>http://arxiv.org/abs/2004.10593</url>
</bibl>

</refgrp>
} 


\section*{Supplementary materials}

\begin{figure}[tbp]
\centering
    \includegraphics[width=0.45\textwidth]{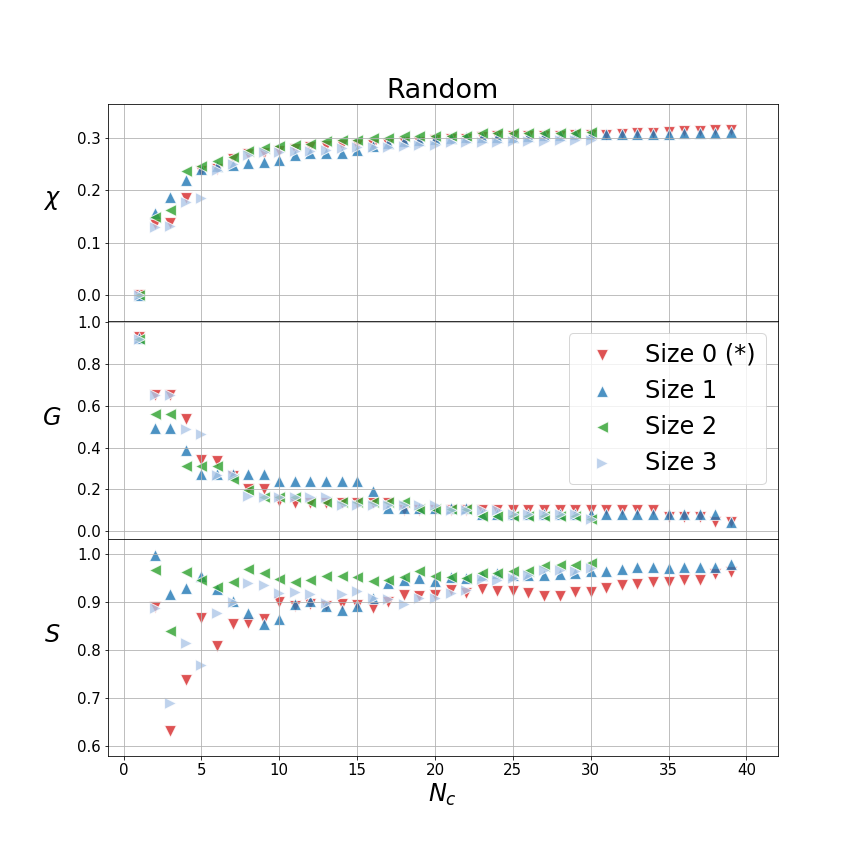}
    \includegraphics[width=0.45\textwidth]{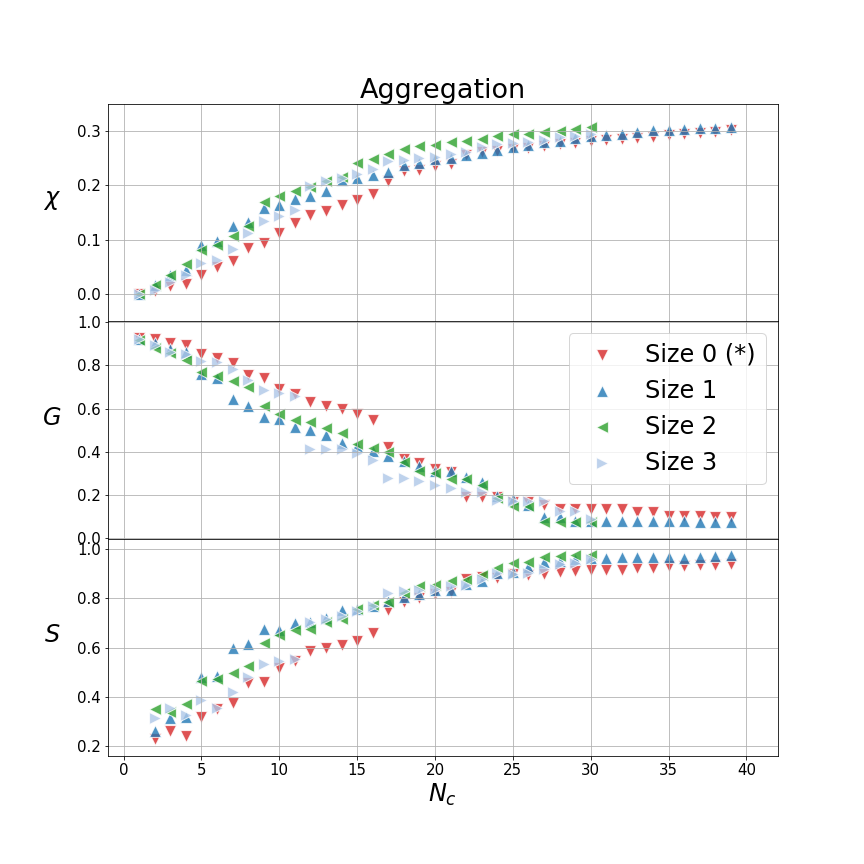}
    \includegraphics[width=0.45\textwidth]{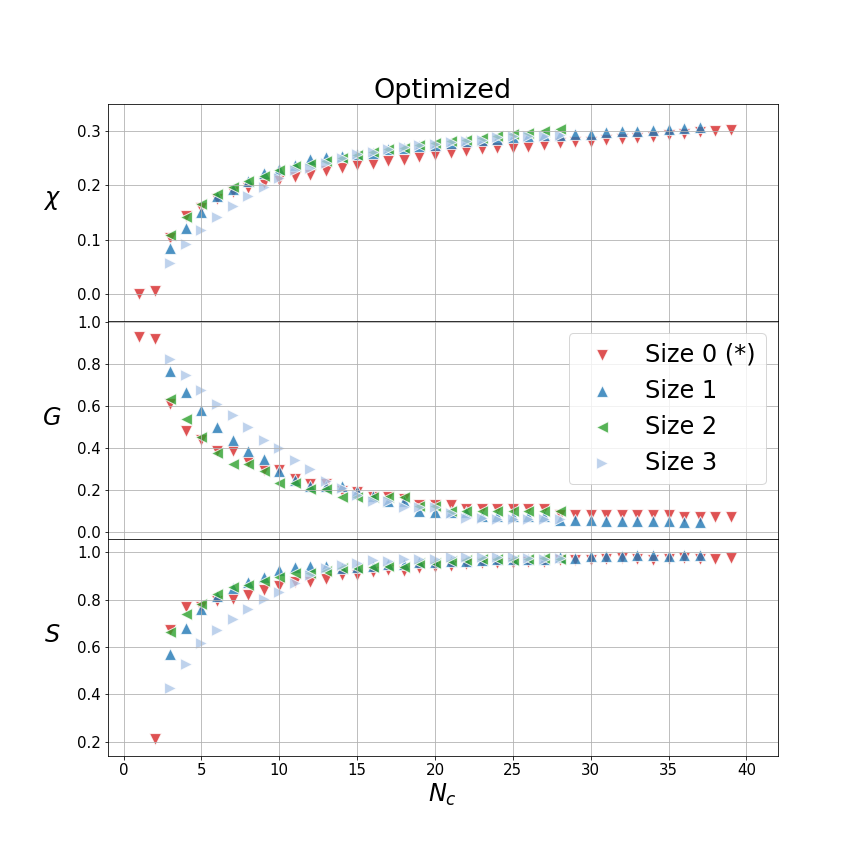}
    \includegraphics[width=0.45\textwidth]{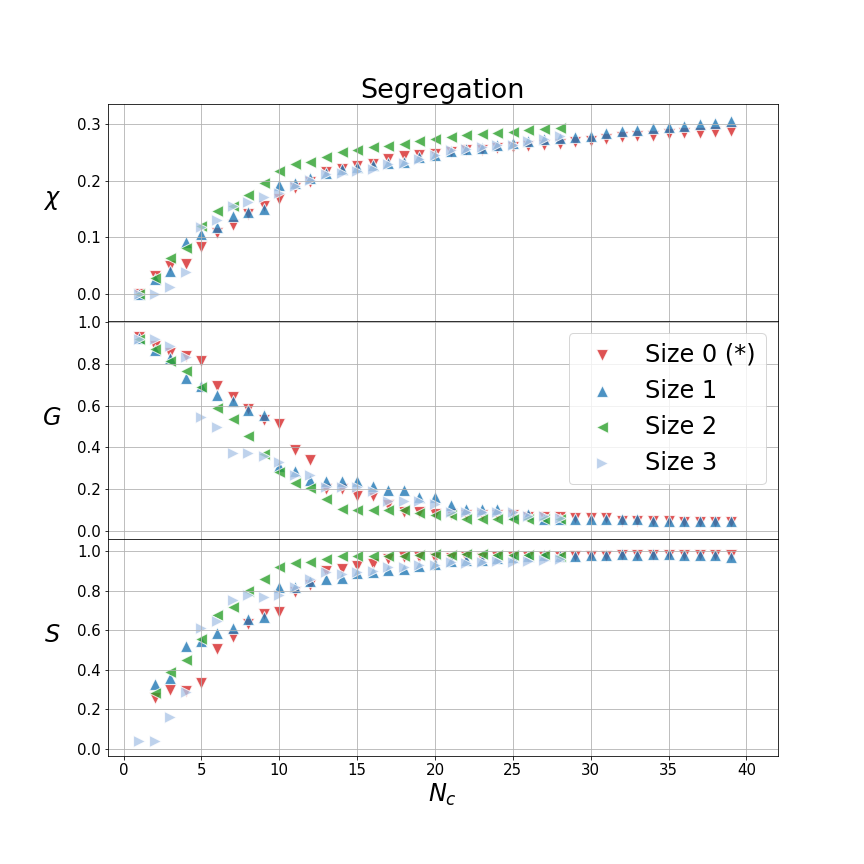}
\caption[short]{Dependence of the results on the workplace size distribution, for the different containment strategies (\emph{Random}, \emph{Aggregation}, \emph{Optimized}, \emph{Segregation}). The number of nodes is $N_n = 1000$. The asterisk (Size 0) refers to the setting that we use in the main analysis of the paper. 
 All the distribution of workplace sizes are Gaussian with different means and standard deviation. 
Size 0: mean 15, std 5; 
Size 1: mean 20, std 5; 
Size 2: mean 25, std 5; 
Size 3: mean 20, std 10.
The influence of the company size distribution is minimal.  
}
\label{fig:company_size}
\end{figure} 

\begin{figure}[tbp]
\centering
    \includegraphics[width=0.45\textwidth]{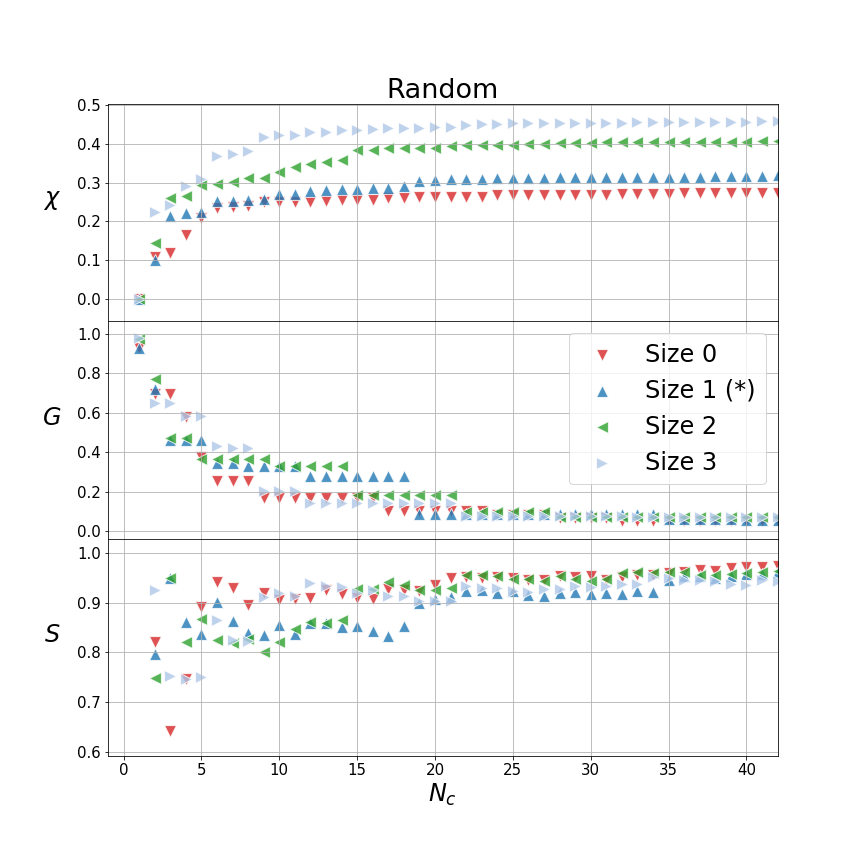}
    \includegraphics[width=0.45\textwidth]{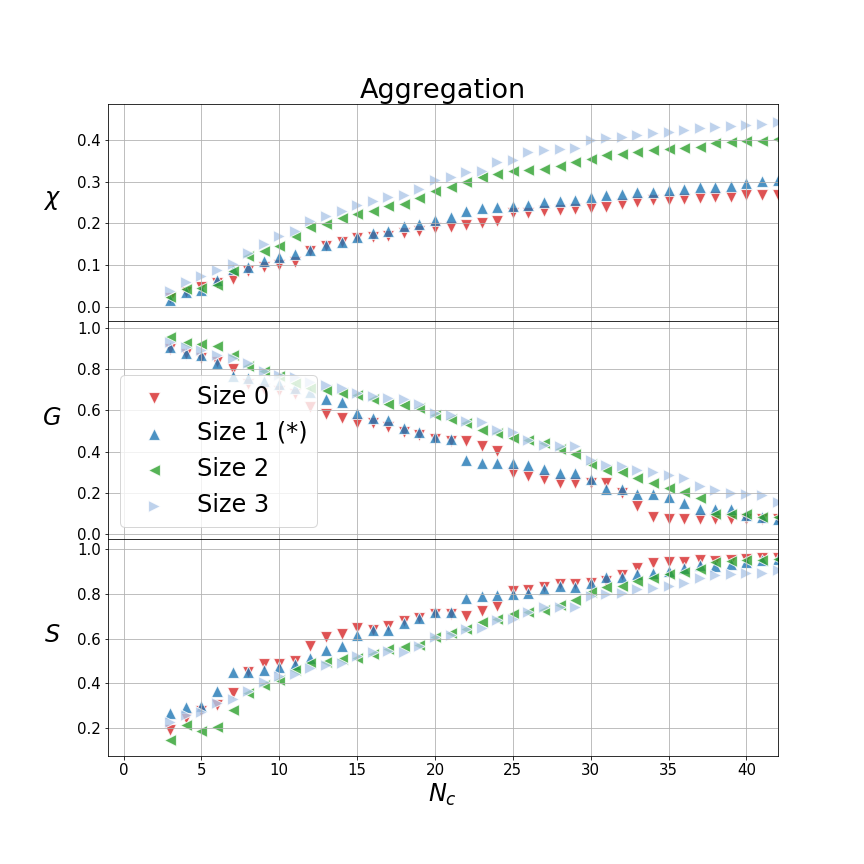}
    \includegraphics[width=0.45\textwidth]{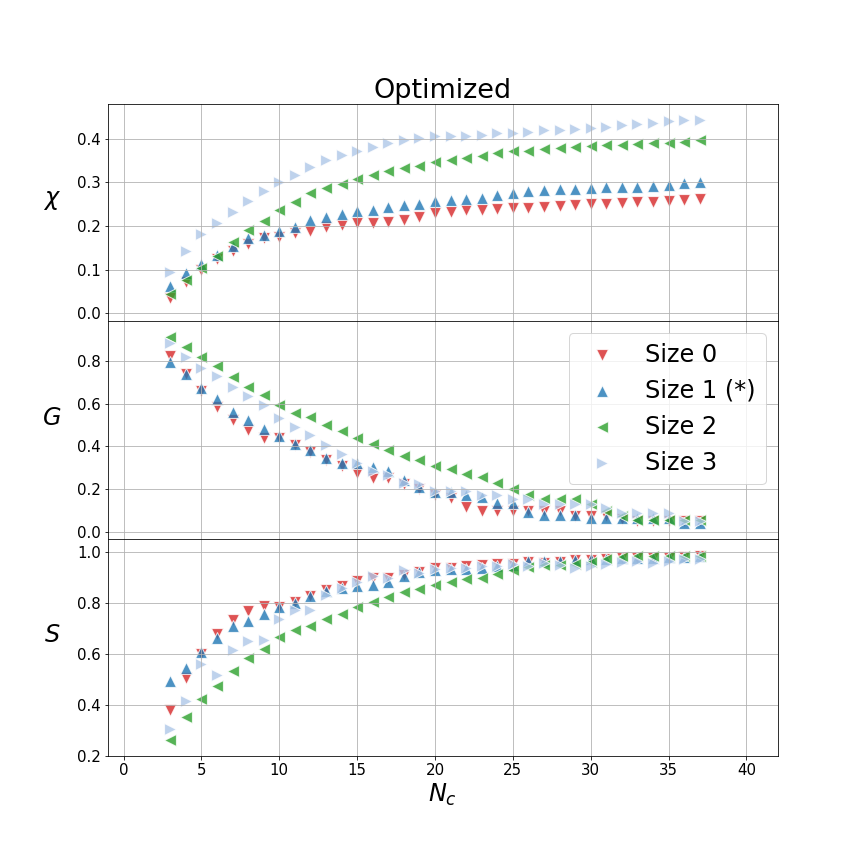}
    \includegraphics[width=0.45\textwidth]{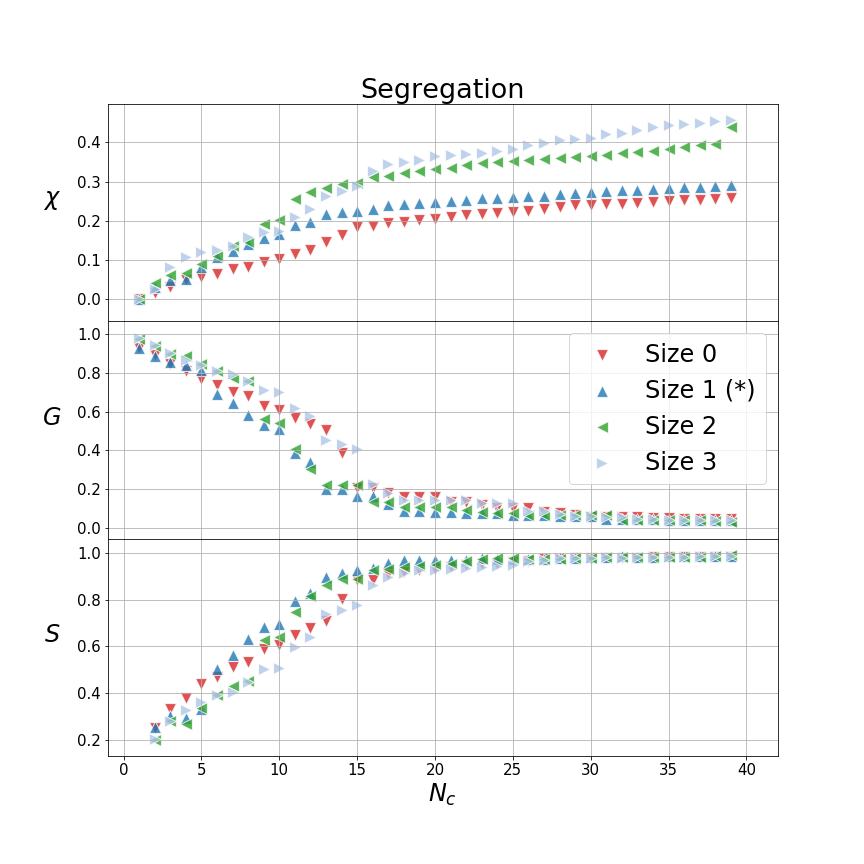}
\caption[short]{Dependence of the results on the household size distribution, for the different containment strategies (\emph{Random}, \emph{Aggregation}, \emph{Optimized}, \emph{Segregation}). The number of nodes is $N_n = 1000$. The asterisk (Size 1) refers to the setting that we use in the main analysis of the paper.  Distribution of household sizes: 
Size 0:  1 member $10\%$, 2 : $80\%$, 3: $10\%$;\\
Size 1: 1 member $38\%$, 2 : $38\%$, 3: $14\%$, 4: $8\%$, 5: $1.5\%$, 6: $0.5\%$; \\
Size 2: 1 member $23\%$, 2 : $28\%$, 3: $19\%$, 4: $13\%$, 5: $11.5\%$, 6: $5.5\%$; \\
Size 3: 1 member $13\%$, 2 : $18\%$, 3: $19\%$, 4: $23\%$, 5: $16.5\%$, 6: $5.5\%$, 7: $5\%$. 
We notice that households with more members  give rise to a higher number of conflicts and bigger connected components. 
}
\label{fig:family_size}
\end{figure}



\end{backmatter}
\end{document}